\documentclass[twocolumn]{aa}  

\usepackage{graphicx}
\usepackage{txfonts}
\usepackage{xcolor}
\usepackage{multirow}
\newcommand{\ME}{M$_{\oplus}$} 
\newcommand{\RE}{R$_{\oplus}$} 
\newcommand{\MS}{M$_{\odot}$} 
\newcommand{\ffHHe}{f$_{\text{H-He}}$}

\begin{document}

   \title{Why do more massive stars host larger planets? }


   \author{M. Lozovsky\thanks{E-mail: michloz@mail.com}
          \inst{1}\fnmsep\inst{4}
          \and
          R. Helled\inst{1}
          \and 
          I.  Pascucci\inst{2}
          \and 
          C. Dorn\inst{1}          
          \and 
          J. Venturini\inst{3}
          \and
          R. Feldmann\inst{1}          
          }

   \institute{Center for Theoretical Astrophysics \& Cosmology \\Institute for Computational Science, University of Zurich, Zurich, Switzerland
         \and
            Department of Planetary Sciences, The University of Arizona, USA
         \and
            International Space Science Institute, Bern, Switzerland    
        \and
            Department of Geosciences, Tel-Aviv University, Tel-Aviv, Israel 
             }

   \date{\today}

 
  \abstract
{}
  {It has been suggested that planetary radii increase with the stellar mass, for planets below 6 R$_{\oplus}$ and host below 1 \MS. 
In this study, we explore whether this inferred relation between planetary size and the host star's mass can be explained by a larger planetary mass among planets orbiting more massive stars, inflation of the planetary radius due to the difference in stellar irradiation, or different planetary compositions and structures. }
   {Using exoplanetary data of planets with measured masses and radii, we investigate the relations between stellar mass and various planetary properties for G- and K- stars, and confirm that  more massive stars host larger planets and more massive. 
   {We find that the differences in the planetary masses and temperatures are insufficient to explain the measured differences in radii between planets surrounding different stellar types. 
    We show that the larger planetary radii can be explained by a larger fraction of volatile material (H-He atmospheres) among planets surrounding more massive} stars.}
   {We conclude that planets around more massive stars are  larger most probably as a result of larger H-He atmospheres. Our findings imply that planets forming around more massive stars tend to accrete H-He atmospheres more efficiently.}


   \maketitle
%

\section{Introduction}

The last couple of decades of exoplanetary exploration has led to the discovery of several thousands of exoplanets. The interior characterization of these planets has become a key objective in exoplanetary science \citep[e.g.,][]{Rogers2010,Dorn2015,unterborn2016scaling,brugger2017constraints,owen2017evaporation,baumeister2020machine,Otegi2020b}. 
We are now in a stage at which exoplanetary characterization is becoming  more detailed, allowing us to understand the various planetary populations within our Galaxy, to make the links to planet formation and evolution models, and to better understand the uniqueness of our own planetary system.
\par

The detected planets with measured radii from the \textit{Kepler} mission \citep{Borucki2010}, whose also have mass determination using  radial velocity (RV) follow-ups or transit timing variations (TTV) can be used to infer the planetary average density, and therefore, to estimate the interior composition. 
By now, we have learned that planets are very diverse astronomical objects, and that they do not follow a single mass-radius (hereafter M-R) relation, but instead, show an intrinsic scatter \citep{Wolfgang2015b,Chen2017,ning2018predicting,Neil2020}.
This scatter suggests a large diversity in the planetary formation and evolution, as well as the interior structure and bulk composition \citep[e.g.,][]{Weiss2014,Turrini2015,Petigura2018,plotnykov2020chemical}.
While determining the planetary structure and composition from measured mass and radius is a highly degenerate problem, because very different compositions and structures can yield similar masses and radii \citep[e.g.,][]{Rogers2010,Dorn2015,Dorn2017}, 
some trends and limits on the planetary compositions can be found using statistical analysis \citep[e.g.,][]{Rogers2015,owen2017evaporation,Lozovsky2018,Neil2020,Otegi2020}.\\

The relatively large database of exoplanets also allows to investigate the dependence of the planetary populations on the stellar mass \citep[e.g.,][]{Fulton2018, Neil2018, Mulders2015, Pascucci2018} and metallicity \citep[e.g.,][]{Mulders2016,Mulders2018,Adibekyan2019,Kutra2020}. 
There are contradictory claims in previous studies on how the {\it Kepler} mini-Neptunes and super-Earths depend on stellar mass. For example, \citet{wu2019mass} argued for a linear relationship between planetary mass and stellar mass, and no correlation between planet radius and host star metallicity. In contrast, \citet{owen2018metallicity} suggested that planetary radii  are correlated to stellar metallicity: Stars of higher metallicity host larger planets.

 Furthermore, \cite{Neil2018} found no strong evidence for host star mass dependence in the M-R relation, using the hierarchical Bayesian modeling  approach presented by  \citet{Wolfgang2016}.

Given the result above, \citet{Pascucci2018} applied \citet{Chen2017} M-R relation to convert the planetary radii measured by \textit{Kepler} into planetary masses. When computing completeness-corrected occurrence rates versus stellar mass, it was found that the mass of the most common planet for relatively small stars (with masses $<$ 1 \MS) scales almost linearly with stellar mass. 
\citet{Pascucci2018} showed that the most common planetary mass depends on the mass of the stellar host. 
More specifically, a  universal (i.e., independent on the stellar type) peak in the occurrence rate of the planet-to-star mass ratio for the \textit{Kepler} planets was found.

\begin{table*}[h]   \caption { Planet samples mean values, divided by stellar types}
	\centering
\begin{tabular}{|l|l|l|l|l|l|l|l|l|}
\hline
Stellar type  & Stellar T (K)      & No. & $\mu$(R)  & $\sigma$(R) & $\mu$(M)  & $\sigma$(M) & $\mu$ (T$_{eq}$)  & $\sigma$(T$_{eq}$)  \\ \hline
F    & 5200-6000 & 8  & 3.32  & 1.84  & 8.03 & 4.17  & 1085.80 & 489.67 \\ \hline
G    & 5200-6000 & 51 & 2.84  & 1.19  & 9.41 & 4.42  & 943.09  & 339.35 \\ \hline
K    & 3700-5200 & 32 & 2.09  & 0.61  & 7.33 & 4.10  & 822.02  & 315.55 \\ \hline
M    & 2400-3700 & 16 & 1.72  & 0.92  & 4.13 & 3.17  & 529.49  & 241.09 \\ \hline
\end{tabular}

  \tiny{Planet samples mean values (as described in section \ref{subsec:Sample}), divided by stellar types. T$_{eq}$ is equilibrium planet temperature that assumes zero albedo (see section \ref{subsec:Teq}). No. stands for number of planets in the sample, $\mu$ stands for mean value and $\sigma$ for standard deviation, assuming normal distribution. Planetary mass, radius and temperatures are given in \ME, \RE~ and K, respectively.  }\label{tab:Stars}
\end{table*}

The study of \citet{Pascucci2018} focuses on MKG stars and suggests that the most common planet around a lower mass stars is smaller in \textit{mass}. For planets in the radius range of 1-6~\RE, \citet{Chen2017} M-R relation are presented by a power-law. Therefore, \citet{Pascucci2018} data indicate that the most common planet around a lower mass stars is smaller in \textit{radius}. 
Most of the planets in the sample used by \citet{Pascucci2018} have radii above 1.6~\RE, and therefore are likely to have some volatile materials. The addition of such low-density materials (even if very small in terms of mass) can lead to much larger radii for small mass planets \citep{Rogers2015,Lozovsky2018}.   M-R relations of volatile-rich planets are more complex and can vary depending on the  planetary temperature and the actual distribution of the material in the interior  \citep[e.g.,][]{Lozovsky2018,Mousis2020,2020Mueller}. In addition, it was recently shown that intermediate-size planets ($\sim$5 \ME -- 20 \ME) can follow two different M-R relations since they can belong to two different planetary populations \citep{Otegi2020}. 

In this work we explore potential reasons why more massive stars tend to host larger planets. In principle, there are at least three possible explanations for the identified correlation between the planetary radius and stellar mass: 
\begin{itemize}
\item Planets are larger due to thermal inflation: since more massive stars are more luminous, larger stellar irradiation could inflate the planetary radius (Case-1). 
\item Planets around more massive stars are more massive and are in consequence larger for any given composition (Case-2). 
\item Planets around more massive stars tend to be more volatile-rich and are therefore larger in size (Case-3). 
\end{itemize}

Below we investigate these three different cases. 
In Case-1, we would expect to find a correlation between planetary equilibrium temperature and radius. Case-2 can be examined by testing whether the observed differences in planetary radius can be explained by a variation of planetary mass, and Case-3 by testing the effect of planetary composition on planetary radius. In this work, we discuss the three cases and find that Case-3 is the favorable interpretation.

\section{Methods}
\label{sec:methods}

In this work we apply a simple statistical analysis of exoplanetary data combined with internal structure models to explore the potential explanations for the observed trend that planets surrounding more massive stars are larger.  

The planetary parameters that are used in the analysis are the planetary mass, radius and insolation flux; the stellar parameters are stellar mass, radius and effective temperature.\\
First, we build an exoplanetary sample based on the \textit{NASA} Exoplanet Archive. The planetary sample is used to quantify the differences between different planetary physical properties, among exoplanets surrounding different stellar populations.

We next construct a series of planetary models using structure equations in order to quantify the effect of different parameters, such as planetary mass, temperature and composition on the planetary radius (section \ref{subsec:MR}) and explore whether  the difference in the observed distributions of equilibrium temperatures (Case-1) and  masses (Case-2) is sufficient to explain the difference in the observed radii. 
Finally, we investigate the effect of various assumed compositions (Case-3) on the inferred radii.

\subsection{Planetary Samples}
\label{subsec:Sample}

\begin{figure}[h]
	\centering

	\includegraphics[width=1\columnwidth, trim={3.8cm 7.7cm 4.3cm 8.6cm},clip]{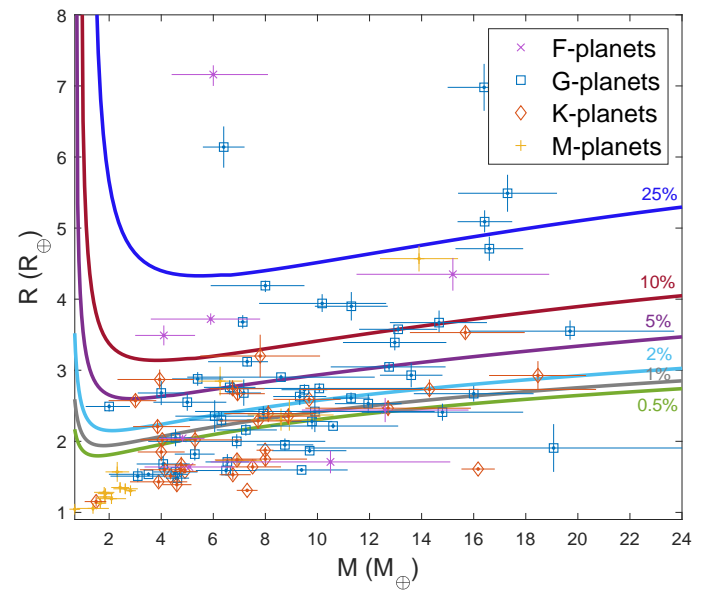}
	\caption{The M-R diagram for planetary sample used in this study. The dots with error bars correspond to F-planets (purple crosses), G-planets (blue squares),  K-planets (red diamonds), and M-planets (yellow pluses). Only G-planets and K-planets are used in the analysis. The colored solid curves are M-R relations for various theoretical models with different mass fractions of H-He (see Section \ref{subsec:MR} for details) and constant H$_2$O mass fraction in the envelope. The presented curves correspond to models with T$_{eq}$ of 700 K (showed as examples).}
	\label{fig:T700_900}
\end{figure}

The planetary sample we use in this study is based on the online publicly available \textit{NASA} Exoplanet Archive.
Since we focus on relatively small planets, we select transiting planets, with \textit{RV} or \textsl{TTV} follow-ups that have masses M$_{pl}$ up to 20 \ME, radii R$_{pl}$ up to 8 \RE~ and uncertainties on mass and radius up to 50\% of the measured value. We divide the planet samples according to stellar type, which is derived from the stellar effective temperatures (see Table \ref{tab:Stars}). 
We determine the average radius, mass, and equilibrium temperature (see section \ref{subsec:Teq} for details) of the planets for each stellar type using the catalogue.

\begin{figure}[h]
	\centering
	\includegraphics[width=1\columnwidth, trim={3.9cm 7.8cm 3.8cm 8.7cm },clip]{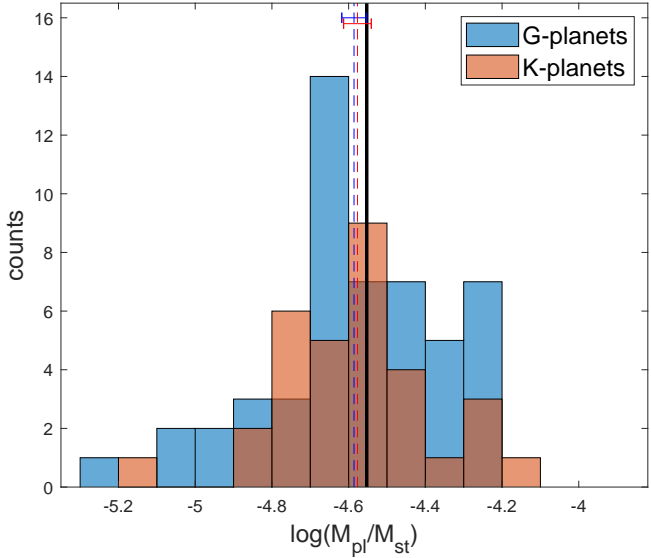}
		\caption{Histogram of planet-to-star mass ratio for the two sub-samples. It can be compared to Figure 1 of Pascucci et al. (2018). The value of q$_{br}$ of \citet{Pascucci2018} of $\log(M_{pl}/M_*)=-4.55\pm0.03$ is presented as a vertical black line and can be compared with mean values we calculated: $-4.59\pm0.03 $ (G-planets, red line) and  $-4.58 \pm 0.04$ (K-planets, blue line). }
	\label{fig:IllariaHist}
\end{figure}

Table \ref{tab:Stars} lists the mean values ($\mu$) and the corresponding standard deviations ($\sigma$) calculated assuming a normal distribution. In this work we consider only planets around K and G stars.  Hereafter, we refer to planets surrounding G and K stars as "G-planets" and "K-planets", respectively. We do not include planets around F stars in our study, because F-planets do not follow the "universal  planet-to-star mass ratio peak" found by \citet{Pascucci2018} and because of the rather small number of F-planets in the planetary sample. M-planets are also not included because of their small number in the planetary database. 
The planets we use are listed in the Appendix.

The database used in this study differs from one used by Pascucci et al.~(2018),  since the \textit{NASA} catalogue is inhomogeneous in terms of detection methods, measurement uncertainties, and selection effects. While Pascucci et al.~(2018) used \textit{Kepler} planets with well-studied data incompleteness, the  \textit{NASA}'s database lists planets detected by various methods, such as Radial Velocity (RV), transit, and direct imagining, and therefore the dataset suffers from different selections effects and data incompleteness. In addition, by selecting planets with measured masses and radii with relatively low uncertainties, we may introduce our own selection. This topic is addressed in Section \ref{subsec:Selection}. The uncertainties of our sample are discussed in Section \ref{sec:uncert}.

Given the potential biases of our planetary catalog, We explore whether the planet-to-star mass ratio inferred by \citet{Pascucci2018} persists when using our sample.
In order to investigate the existence of a universal peak in planet-to-star mass ratio we first calculate the mass ratios for two sub-samples: G-planets and K-planets. The distributions of planet-to-star mass ratio of the two sub-samples is shown in Figure \ref{fig:IllariaHist}, with the mean values (dashed horizontal lines) and corresponding uncertainties of the mean value. This can be compared to the peak value q$_{br}$ from \citet{Pascucci2018} which is given by the solid black line. The figure clearly shows that G-planets and K-planets have a similar mean value, and that this value is in agreement with q$_{br}$ inferred by  \citet{Pascucci2018}. Our results confirms that the mean planetary mass  scales with stellar mass.

It is encouraging that despite the large differences in the data and different selection effects, a similar behavior is found and supports the use of our sample to further investigate the correlation origin.  

\subsubsection{Planetary Equilibrium Temperature}
\label{subsec:Teq}

\begin{figure}
	\centering
		\includegraphics[width=1\columnwidth, trim={3.3cm 5.7cm 3.4cm 5.9cm},clip]{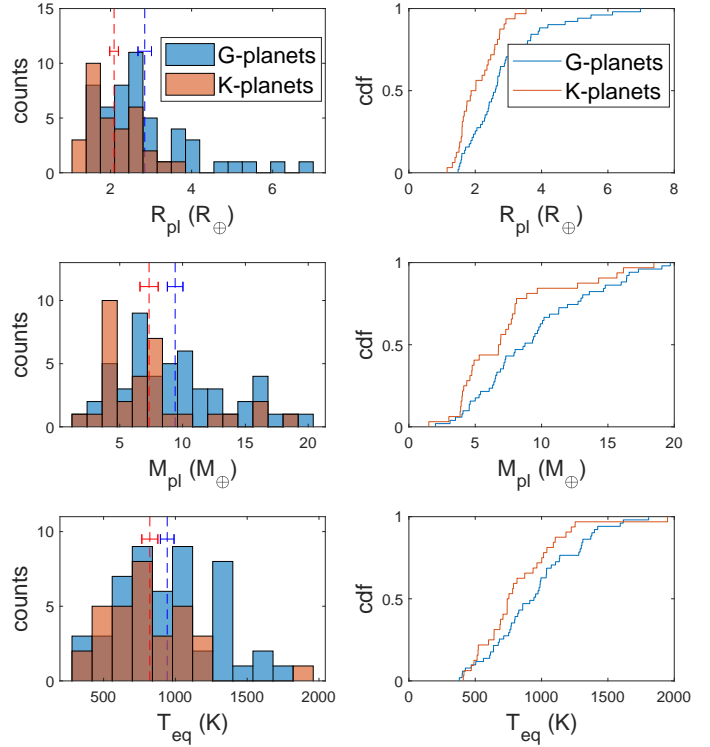}
	\caption{Distributions of planetary radius, mass and equilibrium temperature of G- and K- planets. The left panels show histograms with the mean values and the corresponding uncertainties (dashed lines). The right panels  present the corresponding CDFs. }
	\label{fig:HistoVer2}
\end{figure}

The planetary equilibrium temperatures of the sample are calculated assuming an albedo $A=0$. 
The planetary albedo depends on various parameters, such as atmospheric composition, thermal state, cloud layers etc.; by assuming an albedo of zero, corresponding to full absorption, we infer the highest possible equilibrium temperature. 
The planetary temperature is calculated via:
\begin{equation}
{ T }_{ eq }={ T }_{*}{ \left( 1-A \right)  }^{ 1/4 }\sqrt { \frac { { R }_{*} }{ 2D }  } ,
	\label{eq:Teq1}
\end{equation}
where T$_{ eq }$ is planetary equilibrium temperature, ${ T }_{*}$ is the stellar temperature, $R_{*}$ is the stellar radius, and $D$ is the orbital separation. The values for these parameters are taken from the planetary database. 

Planets with equilibrium temperatures above 2000 K are not considered in this study, because these ultra-hot planets are rare outliers and are likely to be subject to other processes, such as atmospheric loss by photo-evaporation  \citep[e.g.][]{Rosotti2013,Jin2018}, that could affect their basic planetary properties. 
The mean values of T$_{ eq }$, mass and radius of the samples are listed in Table \ref{tab:Stars}, with the corresponding standard deviations. The distributions of radii, masses and equilibrium temperatures are presented in Figure \ref{fig:HistoVer2}. 

We performed a  K-S test on the sub-samples. The test rejects the null hypotheses that G-planets' and K-planets' masses and radii come from the same distribution at the 5\% significance level and asymptotic \textit{p}-values of 0.03 and 0.04,  respectively. The test could not reject the null hypothesis that the equilibrium temperatures of G-planets and K-planets come from the same distribution, with asymptotic \textit{p}-value of 0.18.

\subsubsection{Measurement uncertainties}
\label{sec:uncert}

\begin{figure}[h!]
	\centering

	\includegraphics[width=1\columnwidth, trim={3.5cm 7.8cm 4.1cm 8.3cm},clip]{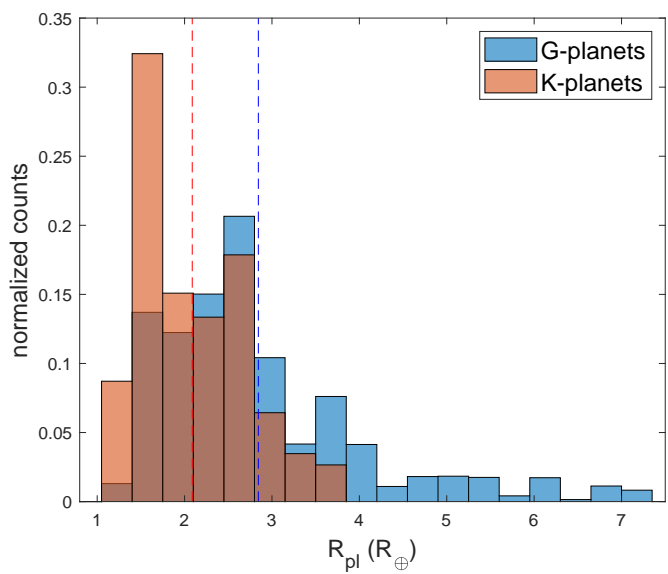}
	\caption{Simulated values of planetary radii accounting for the measurement  uncertainties. For each planet in the sample we simulated 10,000 radii, taking into account the uncertainty. The mean values are presented by the dashed lines. The histogram of the simulated populations show the same tendency as the original one (see Figure \ref{fig:HistoVer2}).} 
	\label{fig:RandomHist}
\end{figure}

The planets we include in the samples have small measurement  uncertainties, and therefore the planetary mass, radius and orbital period are rather well-constrained. In order to examine how the measurement uncertainties affect the main tendency we perform the following test:\\
For each planet in the sample we produce 10,000 synthetic data-points using the measured radii and their uncertainties. The simulated values are probed assuming a normal distribution $R_{pl}\sim \mathcal (R,R_{err})$, where $R$ is the measured radius and $R_{err}$ is the corresponding measurement uncertainty. The simulated values are presented in Figure \ref{fig:RandomHist}. As can be seen from the figure, the main trend of larger stars hosting larger planets persists.  

\subsection{Selection effects}
\label{subsec:Selection}

Most of the detected exoplanets with measured radii were observed with the \textit{Kepler} mission using the transit method, where the transit depth is given by: 
\begin{equation}
 \frac{dF}{F_{st}} = \left( \frac{R_{pl}}{R_{st}} \right)^2. \label{eq:Rratio}
 \end{equation}
Here $dF$ is the dimming due to the transit and $F_{st}$ is the stellar flux \citep[e.g.,][]{Bozza2016}. 
As a result, smaller planets are easier to detect around smaller stars but are easier to miss around larger stars,  
thus, potentially introducing  an artificial correlation between planetary size and stellar size (or stellar mass).

In order to test whether our sample suffers from a strong selection effect of this kind we compare our sample with the  wider (and homogeneous) sample of \textit{Kepler} and \textit{K2} planets, using the \textit{NASA} catalog, without restricting the planetary mass. 
Specifically, we select planets with sizes up to 8 \RE{} and uncertainties below 50\% of the measured values, surrounding G-stars and K-stars with measured stellar radii and temperatures.

\begin{figure}[h!]
	\centering
	
	\includegraphics[width=1\columnwidth, trim={3.5cm 7.8cm 4.3cm 8.0cm},clip]{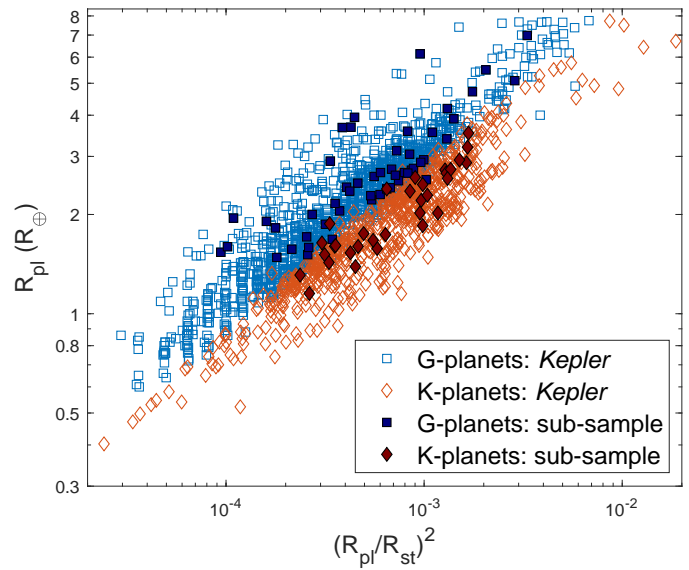}
	\caption{{Planetary radius versus the square of the planet-to-star radius ratio $(R_{pl}/R_{st})^2$ for \textit{Kepler} G-planets (blue squares) and K-planets (red diamonds). The planets shown by filled symbols is the  sub-sample used in this study.}} 
	\label{fig:RratioKepler}
\end{figure}

Figure \ref{fig:RratioKepler} shows the planetary radius as a function of the  planet-to-star radius ratio $(R_{pl}/R_{st})$. 
In this figure, the full sample consisting of 1267 G-planets and 643 K-planets is shown by empty symbols  while the sample we use in this study is shown by the filled symbols. 
In contrast to the expectation from a transit depth related selection bias, neither the full Kepler sample nor our sub-sample show a lack of detected planets below some critical transit depth, i.e., a vertical cut-off in $(R_{pl}/R_{st})^2$. In fact, both G- and K- stars in our sub-sample appear to host  planets as small as  1.2-1.4 \RE{}.
In addition, our sub-sample has planet-to-star radius ratios significantly larger than the minimal value of $\sim{}3\times{}10^{-5}$ in the full sample. This  suggests that selection effects due to small transit depths are unlikely to affect the main results of this study.

\subsection{Interior Modeling}
\label{subsec:MR}
In order to explain the observed differences in planetary properties of different populations, we infer the relation between the planetary mass and radius (M-R relations)  for planets with different assumed compositions and internal structures. It is reasonable to model the planets in the samples with volatile-rich interiors, since most of them have radii larger than the ones expected for an Earth-like composition  \citep[e.g.,][]{Rogers2015,Lozovsky2018}.  
The planetary structure models used consist of a rocky core surrounded by a hydrogen-helium (H-He) and a water (H$_2$O) envelope. 
For simplicity, the composition of the core is set to be pure MgSiO$_3$. More complex core structures are expected to have a negligible  effect on the M-R relation for volatile-rich planets \citep{Howe2014,Otegi2020}. 
The H to He ratio is assumed to be protosolar, i.e., 72\% H to 28\% He by mass, and the water mass  fraction is taken to be 20\% in the  envelope (water + H-He). The equation of state for H-He is taken from \citet{Saumon1995}, for MgSiO$_3$ from \citet{Seager2007} and for H$_2$O from  \citet{Vazan2013}. 
The atmospheric models are derived using the standard structural equations of hydrostatic equilibrium, mass conservation, and heat transport. We follow the irradiation model of \citet{Guillot2010} for the H-He envelope. Further details on the structure models can be found in \citet{Venturini2015}, \citet{Dorn2017}, \citet{Lozovsky2018} and references therein. 
We consider two cases of envelope structure: a homogeneously mixed envelope consisting of H-He and water and a differentiated structure, where the water and H-He are separated, and the pure water layer is below pure H-He layer. The first case (homogeneous envelope) is used as default. 

The structure models are constructed assuming various percentages of H-He between 0.5\% and 25\% of the total planetary mass and different equilibrium temperatures T$_{eq}$ (see section \ref{subsec:Teq} for details). Models with H-He mass fractions larger than 25\% were not considered since we focus on planets up to 20 \ME, which are unlikely to be H-He-rich. 
The M-R relations, together with the  planetary sample are shown in Figure \ref{fig:T700_900}; the curves presented in this plot are examples assuming T$_{eq}$=700 K. The calculated M-R relations allow us to derive the planetary radius corresponding to any given mass and T$_{eq}$. These models are used in the following analysis in order to investigate the potential reason(s) for the trend of "larger planets around more massive stars". 

\section{Results}

\subsection{Case-1: Larger Planets due to Inflation}
\label{subsec:TempEffect}

\begin{figure}[h!]
	\centering
		\includegraphics[width=1\columnwidth, trim={4.0cm 7.7cm 4.2cm 5.1cm},clip]{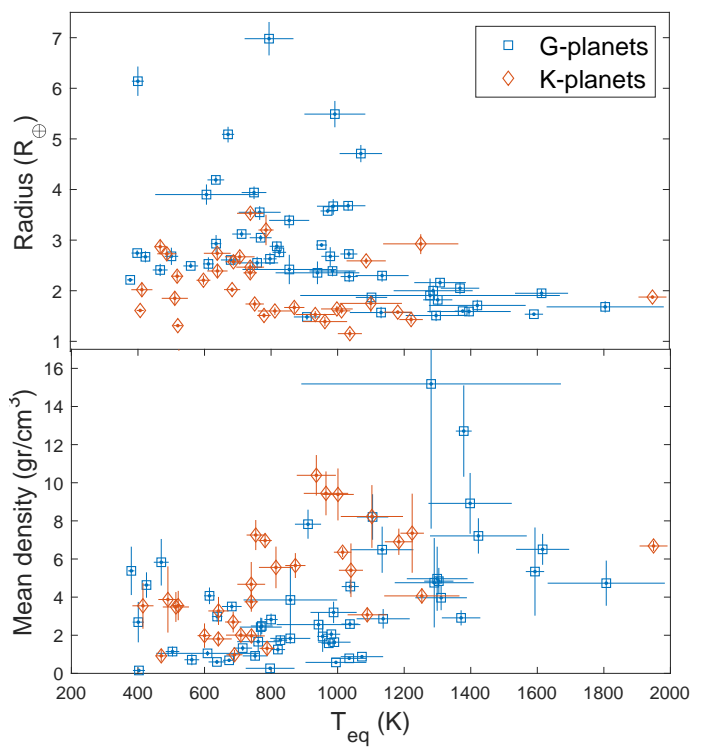}
	\caption{Measured planetary radius (upper panel) and calculated planetary bulk density (lower panel) versus calculated equilibrium temperature for our sample of planets.}
	\label{fig:RvcT}
\end{figure}

As mentioned above, most of the planets in the sample are expected  to be volatile-rich given their large radii and low bulk densities (Figure \ref{fig:RvcT}). 
Radii of volatile-rich planets are very sensitive to the temperature: planets with the same mass,  chemical composition, and internal structure can have different radii for different temperatures due to thermal inflation. Planets surrounding more massive stars seem to be slightly hotter on average (see Figure \ref{fig:HistoVer2}). Therefore, it is reasonable to expect that the differences in the equilibrium temperatures for G- and K-planets could lead to differences in radii of the two populations. Below, we use the planet sample to evaluate any possible correlation between planetary equilibrium temperatures and radii.

Figure \ref{fig:RvcT} (upper panel) presents the observed planetary radii versus equilibrium temperature T$_{eq}$, calculated from Equation \ref{eq:Teq1}. Interestingly, the  observed radii do not increase with temperature. G- and K-planets even tend to have smaller radii in for higher temperatures.
Figure \ref{fig:RvcT} (lower panel) presents the bulk densities (inferred  from the measured masses and radii) versus T$_{eq}$ for the detected planets. No significant decrease of the bulk density with temperature is observed. In fact, the  bulk densities of both G- and K-planets is increasing with temperature, suggesting that the very hot planets tend to be more rocky. This correlation could be 
a result of atmospheric mass loss of primordial hydrogen-atmospheres, that depends on stellar insolation \citep{Fulton2018, Owen2017, Jin2018}.

Thus, the observed radii do not correlate with the temperature, and thermal inflation does not play a significant role in observed trend. We therefore conclude that the \textit{differences in the equilibrium temperature alone is insufficient to explain the differences in the observed radii of the two planetary populations.}

\subsection{Case-2:  Larger Planets due to Higher Planetary Mass}
\label{subsec:MassEffect}

Assuming that the planetary radius scales with mass, it is reasonable to expect that the observed difference in radii is a result of difference in planetary mass. 
In this section we investigate whether the observed difference in average radii is caused by more massive planets around more massive stars. 

To investigate this option, we use the theoretical M-R relations of planets with rocky cores surrounded by H-He + H$_2$O atmospheres, calculated from the internal structure models described in section \ref{subsec:MR}. The typical difference between masses of G-planets and K-planets in the sample 
is $\sim$ 2 \ME.
The mass difference is calculated as a difference between the  mean values of the distributions. This  difference in mass distributions (see Figure \ref{fig:HistoVer2} and Table \ref{tab:Stars}) is insufficient to explain the difference in the observed radii of 0.75 \RE~, as 
 the M-R curves are very flat in this regime \citep{,Lopez2014,Weiss2013,Wolfgang2015b, Chen2017,Lozovsky2018}, already implying the weak dependency  of the radius for these planetary masses. 

In order to test whether the difference in radii is a result of different planetary masses, we calculate the radius corresponding to a hypothetical average mass planet of each stellar type:
9.41 \ME~ for  G-planets and 7.33 \ME~ for K-planets, as listed in Table \ref{tab:Stars}.
To isolate the effect of mass, we assume a uniform temperature of T$_{eq}$= 900K and H-He mass fraction of 2\% for both hypothetical planets. This method compares between the mean values of the distributions of masses, assuming the average masses to be representative of the main trend; by calculating a radius corresponding to each average mass and using the same M-R relation and temperature for all of the masses, we are able to isolate the influence of the mass alone. The calculated radii of both G-planets and K-planets are found to be in the range of 2.62-2.76 \RE~  for G-planets and 2.58-2.81 \RE~.

These radii ranges are calculated assuming two different atmospheric structures: The lower radius boundary is calculated assuming a homogeneous envelope structure (water and H-He are mixed), while the upper limit is calculated assuming a differentiated structure, where the water and H-He are separated (see section \ref{subsec:MR}).  We find that the effect of the mass on the calculated radius alone is negligible compared to the observed differences.

Therefore, we conclude that the \textit{difference in the planetary masses cannot explain the measured difference in planetary radii.}

\subsection{Case-3: Larger Planets due to Different Compositions}
\label{subsec:CompEffect}

\begin{figure}[h]
	\centering
		\includegraphics[width=\columnwidth, trim={3.8cm 7.8cm 3.9cm 8.6cm},clip]{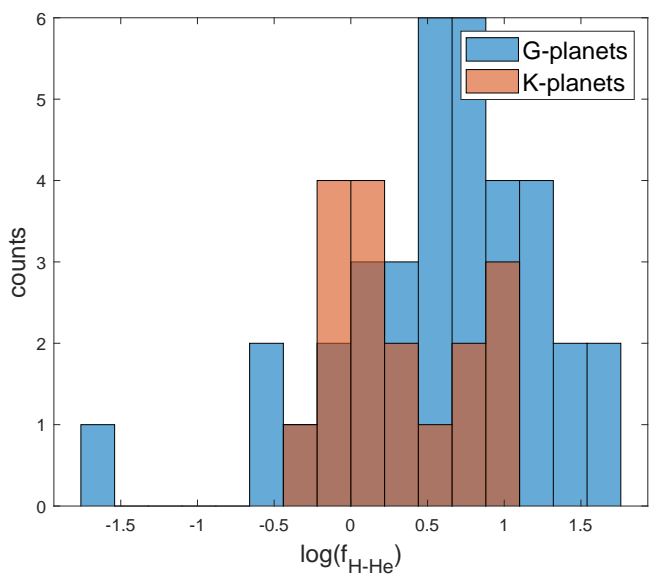}
	\caption{Histogram of simulated \ffHHe~ for our planetary samples. Planets that cannot be modeled with any H-He mass fraction are not presented. }
	\label{fig:fHist}
\end{figure}

Finally, we investigate the effect of different compositions and envelope structures on the planetary radius.  
For planets with thick H-He envelopes, the planetary structure and exact H-He mass fraction have significant effects on the M-R relation (see  \citet{Lozovsky2018} for details). 
For the same planetary mass and temperature the difference between a planet with \ffHHe~ of 0.5\% and \ffHHe~ of 10\% is between $\sim 1.2$ \RE~  and $\sim 1.8$ \RE. This range exceeds the difference between the measured radii of G-planets and K-planets: the difference between the peaks of the radii distribution is 0.75 \RE~.

We next compare the radii inferred when assuming various compositions and internal structures with the measured ones.  Using the planetary models described in section  \ref{subsec:MR}, we calculate for each planet the corresponding \ffHHe. The calculation uses the  planetary effective temperature and the measured mass. The results are shown in Figure \ref{fig:fHist}. 
There is a clear trend where planets around larger stars have higher \ffHHe values: G-planets: $0\lesssim $ \ffHHe $ \lesssim 0.4$, K-planets:  $0 \lesssim $ \ffHHe $\lesssim 0.15$) , with mean values of 0.06$\pm$0.01 and 0.03$\pm$0.01 respectively.
It should be noted that there are several small planets which are unlikely to consist of H-He: this corresponds to 15 G-planets and 15 K-planets.
These planets are probably rocky-water dominated planets without any significant volatile atmosphere, and therefore do not have a strong dependency on T$_{eq}$, and thus do not influence our conclusions. Note that those planets are not presented in Figure \ref{fig:fHist}, as they are unlikely to possess any H-He envelopes and can be seen from Figure \ref{fig:T700_900}, below the 0.5\% H-He  line.

As expected, the H-He mass fraction in the planet \ffHHe~ has a crucial effect on the inferred radii (e.g Figure \ref{fig:T700_900}; \citet{Swift2012,Weiss2014}). The difference in the assumed \ffHHe~ leads to a radius difference that can exceed the difference between G-planets and K-planets. We therefore suggest that \textit{ the difference in observed radii of planets orbiting stars with difference masses  could be caused by different H-He mass fractions, with the G-planets being more volatile-rich in comparison to K-planets}. In addition, it should be noted that the distribution of volatiles has a significant effect on M-R relations: the size  difference between fully differentiated and homogeneous envelopes is larger than the difference due to thermal inflation and larger mass  \citep[e.g.,][]{Soubiran2015}.  

We conclude that while case-1 (thermal inflation) and case-2 (higher mass) cannot explain the observed differences, case-3 (different planetary composition and structure) can explain the observations.

\section{Discussion}
In this study we investigate the potential explanations for the observed correlation between mean planetary size and a stellar type. We show that different H-He mass fractions could explain this observations. 
In the analysis, we restricted the possible theoretical models to a rather narrow range: rocky cores surrounded by H-He+H$_2$O atmospheres, with \ffHHe$=0.5\%-25\%$ and water fraction in the envelope of 20\% by mass. Nevertheless, even in this narrow assumption range, we could provide an interpretation for the reason that the most typical planetary radius scales with stellar mass.  
We suggest that this correlation might be a result of characteristically different planetary composition and/or structure among planets surrounding different stellar types, while planetary mass and insolation are unlikely to play a big role in this relation. Planet formation and evolution models could confirm this idea. In addition, it is unlikely that the found correlation is caused by data incompleteness. Nevertheless, the planetary data used in this analysis are somewhat limited and inhomogeneous. As a result, it is  desirable to use larger samples in the future, and repeat the analysis accounting for a larger range in stellar spectral types, also including M-dwarfs. 

Previous studies have investigated differences in planet populations between M-dwarfs and FGK- dwarfs. Giant planets are scarce around M-dwarfs \citep{Bonfils2013} and there is a higher occurrence of small planets around M-dwarfs \citep{DressingCharbonneau2015,Gaidos2016}. The higher planet-formation efficiency of low-mass stars \citep{dai2020california} may be in fact related to the small giant planet occurrence, or the longer disk lifetimes, or other factors. How the planetary composition distribution differs is yet unclear. So far, there is no strong evidence for a dependency between the planetary M-R relation and the host star's mass for small planets \citep{Neil2018}. 
Our findings, if extrapolated to M-dwarfs, would predict a tendency of H-He-poor planets compared to those around larger stars. This is somewhat in line with previous studies \citep{Bonfils2013,DressingCharbonneau2015,Gaidos2016} as planets around M-stars are less massive. This does not imply that planets are poor in water. In fact, planets around M-dwarfs can be water-rich due to an efficient inwards migration {\citep{alibert2017,Burn2021}}.

{\it Kepler} planets are located relatively close to their stars, and may thus be affected by planet–star interactions; either by stellar insolation during planet evolution or during formation in the protoplanetary disk. Here, we are arguing that stellar insolation alone cannot make up for the different radii in the observed planetary populations around G- and K-stars. Instead we provide evidences that the populations are different due to formation processes, i.e. the efficiency of envelope accretion. This can be a result of more massive protoplanetary disks \citep{mohanty2013protoplanetary,andrews2013mass,pascucci2016steeper,ansdell2016alma} that lead to faster growth \citep{hartmann2006,alcala2014} followed by H-He accretion (even if in small amounts) and/or due to different disk lifetimes \citep{ribas2015protoplanetary}. Essentially, core growth is more efficient both in higher-mass disks {\citep{mordasini2012,Venturini2020}} and around higher-mass stars \citep{Kennedy2008} allowing them to to accrete substantial gaseous envelope before the gas disk dissipates.

The diversity of planetary envelopes of our sample is yet unknown, and different compositions including H, He, water, and higher mean molecular species are possible. Besides compositional diversity, these planets may absorb stellar energy differently which affects their thermal structure and thus their radii.

In that regard, it would  be desirable to investigate the dependence on the assumed albedo; In this work we assumed a uniform zero albedo for all planets when calculating the T$_{eq}$.
Since different compositions can lead to different albedo, the relation between the different parameters is likely to be very complex. 
Another important property that could influence the planetary population is the age of the system. In order to investigate a correlation between the age and size of the planets, we need better constraints on the stellar ages, which is not available for most of the stars in current planetary databases.

{It should be noted that planets around F-stars are not included in this analysis because they do not present the same tendency as G-planets and K-planets, as was shown by \citet{Pascucci2018}. F-stars are more luminous and as a result have higher irradiation. As a result, close-in planets might suffer from other physical effects, such as atmospheric loss by photo-evaporation  \citep[e.g.,][]{Rosotti2013,Jin2018}, that could affect their basic planetary properties and alter the statistics. It is therefore clear that more data and as well as theoretical models that account for the planetary evolution are required before a conclusion can be made regarding the expected compositions of planets around F-stars.}

\section{Summary and Conclusions}

We confirm that planets surrounding larger stars tend to be larger in {mass and} radius. In this study we investigated the influence of the planetary mass, equilibrium temperature, and possible bulk composition on the planetary radius for planets surrounding G- and K-stars.
The larger radii of planets surrounding more massive stars cannot be explained by inflation due to higher irradiation of the star, nor by a higher planetary mass: the effect of both properties on the radius was found to be significantly smaller than the observed difference. In other words, we show  that the difference in planetary mass, which is given by \textit{RV/TTV} data, is  insufficient to explain the observed difference in planetary radii{, and the planetary temperature is even found to anti-correlate with the radius}.

Larger H-He mass fractions on planets around larger stars can easily explain the trends in the observed data.\textit{ We therefore suggest that the larger radii of planets surrounding larger stars are a result of different compositions: larger stars tend to host planets with larger H-He mass fractions.}
If this is correct, it provides important constraints for planet formation models, since it implies  that planets forming around more massive stars tend to accrete H-He atmospheres more efficiently. This can be a result of more massive protoplanetary disks that lead to faster core growth; allowing the cores to accrete  substantial gaseous envelopes before the gas disk dissipates. Thus, we argue that planets around G- and K-stars are different by formation.

In that light, it is desirable to have further constraints available to test our findings. Further information on the planetary composition would come from \textit{JWST} and Ariel missions with constraints on atmospheric composition, which will greatly shed more light into the diversity of planetary volatile envelopes, their formation efficiencies and evolutions. In addition, an  important property of the planetary system is the planetary age, which is typically not very well constrained \citep[e.g.,][]{Saffe2005,Aguirre2015,Guillot2014}. This is expected to change with the PLATO mission which will allow to estimate the stellar ages with accuracies of about 10\% \citep{Rauer2014}. With age constraints, we will be able to check if and how the discussed differences in planet populations are evolving in time.

\begin{acknowledgements}
R.H.~acknowledges support from SNSF grant 200021\_169054. C.D. acknowledges support from the Swiss National Science Foundation under grant PZ00P2\_174028. This work has been carried out within the frame of the National Center for Competence in Research PlanetS supported by the SNSF Swiss National Foundation. 
      
\end{acknowledgements}

%
  \bibliographystyle{aa} 
 \bibliography{references.bib} 
%
\newpage
\begin{appendix}

\section{Planetary Data Tables}
\label{app:tables}
\begin{table*}[h]\caption{Planets used in the study}
\renewcommand{\arraystretch}{1.5}
\begin{tabular}{|l|l|l|l|l|l|l|}
\hline
Name             & Mass(M$_\oplus$)       & Radius(R$_\oplus$)    & T$_{eq}$ (K)            & Host Mass(M$_\odot$)  & Host Radius(R$_\odot$) & Host type \\ \hline
CoRoT-7 b        &  4.08$^{+1.02}_{-1.02}$ & 1.68$^{+0.11}_{-0.11}$ & 1807   & 0.82$^{+0.17}_{-0.17}$ & 0.83$^{+0.04}_{-0.04}$ & G    \\ \hline
EPIC 249893012 b &  8.75$^{+1.09}_{-1.08}$ & 1.95$^{+0.09}_{-0.08}$ & 1616   & 1.05$^{+0.05}_{-0.05}$ & 1.71$^{+0.04}_{-0.04}$ & G    \\ \hline
EPIC 249893012 c & 14.67$^{+1.84}_{-1.89}$ & 3.67$^{+0.17}_{-0.14}$ & 990    & 1.05$^{+0.05}_{-0.05}$ & 1.71$^{+0.04}_{-0.04}$ & G    \\ \hline
EPIC 249893012 d & 10.18$^{+2.46}_{-2.42}$ & 3.94$^{+0.13}_{-0.12}$ & 752    & 1.05$^{+0.05}_{-0.05}$ & 1.71$^{+0.04}_{-0.04}$ & G    \\ \hline
HD 110113 b      &  4.55$^{+0.62}_{-0.62}$ & 2.05$^{+0.12}_{-0.12}$ & 1371   & 1.00$^{+0.06}_{-0.06}$    & 0.97$^{+0.02}_{-0.02}$ & G    \\ \hline
HD 136352 b      &  4.62$^{+0.45}_{-0.44}$ & 1.48$^{+0.06}_{-0.06}$ & 911    & 0.91$^{+0.06}_{-0.05}$ & 1.01$^{+0.02}_{-0.02}$ & G    \\ \hline
HD 136352 c      & 11.29$^{+0.73}_{-0.69}$ & 2.61$^{+0.08}_{-0.08}$ & 682    & 0.91$^{+0.06}_{-0.05}$ & 1.01$^{+0.02}_{-0.02}$ & G    \\ \hline
HD 183579 b      & 19.70$^{+4.00}_{-3.90}$ & 3.55$^{+0.15}_{-0.12}$ & 769    & 1.03$^{+0.03}_{-0.03}$ & 0.98$^{+0.04}_{-0.03}$ & G    \\ \hline
HD 219666 b      & 16.60$^{+1.30}_{-1.30}$ & 4.71$^{+0.17}_{-0.17}$ & 1073   & 0.92$^{+0.03}_{-0.03}$ & 1.03$^{+0.03}_{-0.03}$ & G    \\ \hline
HD 86226 c       &  7.25$^{+1.19}_{-1.12}$ & 2.16$^{+0.08}_{-0.08}$ & 1311   & 1.02$^{+0.06}_{-0.07}$ & 1.05$^{+0.03}_{-0.03}$ & G    \\ \hline
K2-111 b         &  5.29$^{+0.76}_{-0.77}$ & 1.82$^{+0.11}_{-0.09}$ & 1304   & 0.84$^{+0.02}_{-0.02}$ & 1.25$^{+0.02}_{-0.02}$ & G    \\ \hline
K2-138 b         &  3.10$^{+1.05}_{-1.05}$ & 1.51$^{+0.11}_{-0.08}$ & 1299   & 0.94$^{+0.02}_{-0.02}$ & 0.86$^{+0.03}_{-0.02}$ & G    \\ \hline
K2-138 c         &  6.31$^{+1.13}_{-1.23}$ & 2.30$^{+0.12}_{-0.09}$ & 1137   & 0.94$^{+0.02}_{-0.02}$ & 0.86$^{+0.03}_{-0.02}$ & G    \\ \hline
K2-138 d         &  7.92$^{+1.39}_{-1.35}$ & 2.39$^{+0.10}_{-0.08}$ & 988    & 0.94$^{+0.02}_{-0.02}$ & 0.86$^{+0.03}_{-0.02}$ & G    \\ \hline
K2-138 e         & 12.97$^{+1.98}_{-1.99}$ & 3.39$^{+0.16}_{-0.11}$ & 858    & 0.94$^{+0.02}_{-0.02}$ & 0.86$^{+0.03}_{-0.02}$ & G    \\ \hline
K2-263 b         & 14.80$^{+3.10}_{-3.10}$ & 2.41$^{+0.12}_{-0.12}$ & 470    & 0.88$^{+0.03}_{-0.03}$ & 0.85$^{+0.02}_{-0.02}$ & G    \\ \hline
K2-265 b         &  6.54$^{+0.84}_{-0.84}$ & 1.71$^{+0.11}_{-0.11}$ & 1423   & 0.92$^{+0.02}_{-0.02}$ & 0.98$^{+0.05}_{-0.05}$ & G    \\ \hline
K2-291 b         &  6.49$^{+1.16}_{-1.16}$ & 1.59$^{+0.10}_{-0.07}$ & 1398   & 0.93$^{+0.04}_{-0.04}$ & 0.90$^{+0.04}_{-0.03}$  & G    \\ \hline
K2-38 b          & 19.07$^{+9.53}_{-9.53}$ & 1.91$^{+0.34}_{-0.34}$ & 1281   & 2.24$^{+1.67}_{-1.67}$ & 1.38$^{+0.21}_{-0.21}$ & G    \\ \hline
K2-38 c          &  9.90$^{+4.60}_{-4.60}$ & 2.42$^{+0.29}_{-0.29}$ & 858    & 1.07$^{+0.05}_{-0.05}$ & 1.1$^{+0.09}_{-0.09}$  & G    \\ \hline
Kepler-107 b     &  3.51$^{+1.52}_{-1.52}$ & 1.54$^{+0.03}_{-0.03}$ & 1593   & 1.24$^{+0.03}_{-0.03}$ & 1.45$^{+0.01}_{-0.01}$ & G    \\ \hline
Kepler-107 c     &  9.39$^{+1.77}_{-1.77}$ & 1.60$^{+0.03}_{-0.03}$ & 1379   & 1.24$^{+0.03}_{-0.03}$ & 1.45$^{+0.01}_{-0.01}$ & G    \\ \hline
Kepler-107 e     &  8.60$^{+3.60}_{-3.60}$ & 2.90$^{+0.04}_{-0.04}$ & 955    & 1.24$^{+0.03}_{-0.03}$ & 1.45$^{+0.01}_{-0.01}$ & G    \\ \hline
Kepler-11 d      &  7.30$^{+0.80}_{-1.50}$ & 3.12$^{+0.06}_{-0.07}$ & 715    & 0.96$^{+0.03}_{-0.03}$ & 1.06$^{+0.02}_{-0.02}$ & G    \\ \hline
Kepler-11 e      &  8.00$^{+1.50}_{-2.10}$ & 4.19$^{+0.07}_{-0.09}$ & 637    & 0.96$^{+0.03}_{-0.03}$ & 1.06$^{+0.02}_{-0.02}$ & G    \\ \hline
Kepler-11 f      &  2.00$^{+0.80}_{-0.90}$ & 2.49$^{+0.04}_{-0.07}$ & 562    & 0.96$^{+0.03}_{-0.03}$ & 1.06$^{+0.02}_{-0.02}$ & G    \\ \hline
Kepler-18 b      &  6.90$^{+3.40}_{-3.40}$ & 2.00$^{+0.10}_{-0.10}$ & 1290   & 0.97$^{+0.04}_{-0.04}$ & 1.11$^{+0.05}_{-0.05}$ & G    \\ \hline
Kepler-18 c      & 17.30$^{+1.90}_{-1.90}$ & 5.49$^{+0.26}_{-0.26}$ & 995    & 0.97$^{+0.04}_{-0.04}$ & 1.11$^{+0.05}_{-0.05}$ & G    \\ \hline
Kepler-18 d      & 16.40$^{+1.40}_{-1.40}$ & 6.98$^{+0.33}_{-0.33}$ & 797    & 0.97$^{+0.04}_{-0.04}$ & 1.11$^{+0.05}_{-0.05}$ & G    \\ \hline
Kepler-20 b      &  9.70$^{+1.41}_{-1.44}$ & 1.87$^{+0.07}_{-0.03}$ & 1105   & 0.95$^{+0.05}_{-0.05}$ & 0.96$^{+0.02}_{-0.02}$ & G    \\ \hline
Kepler-20 c      & 12.75$^{+2.17}_{-2.24}$ & 3.05$^{+0.06}_{-0.06}$ & 772    & 0.95$^{+0.05}_{-0.05}$ & 0.96$^{+0.02}_{-0.02}$ & G    \\ \hline
Kepler-20 d      & 10.07$^{+3.97}_{-3.70}$ & 2.74$^{+0.07}_{-0.06}$ & 401    & 0.95$^{+0.05}_{-0.05}$ & 0.96$^{+0.02}_{-0.02}$ & G    \\ \hline
Kepler-289 d     &  4.00$^{+0.90}_{-0.90}$ & 2.68$^{+0.17}_{-0.17}$ & 504    & 1.08$^{+0.02}_{-0.02}$ & 1.00$^{+0.02}_{-0.02}$    & G    \\ \hline
Kepler-29 b      &  5.00$^{+1.50}_{-1.30}$ & 2.55$^{+0.12}_{-0.12}$ & 761    & 0.76$^{+0.02}_{-0.03}$ & 0.73$^{+0.03}_{-0.03}$ & G    \\ \hline
Kepler-30 b      & 11.30$^{+1.40}_{-1.40}$ & 3.90$^{+0.20}_{-0.20}$ & 609    & 0.99$^{+0.08}_{-0.08}$ & 0.95$^{+0.12}_{-0.12}$ & G    \\ \hline
Kepler-36 c      &  7.13$^{+0.18}_{-0.18}$ & 3.68$^{+0.10}_{-0.09}$ & 1035   & 1.03$^{+0.02}_{-0.02}$ & 1.63$^{+0.04}_{-0.04}$ & G    \\ \hline
Kepler-454 b     &  6.05$^{+1.51}_{-1.51}$ & 2.35$^{+0.22}_{-0.22}$ & 943    & 0.85$^{+0.22}_{-0.22}$ & 1.05$^{+0.07}_{-0.07}$ & G    \\ \hline
Kepler-538 b     & 10.60$^{+2.50}_{-2.40}$ & 2.21$^{+0.04}_{-0.03}$ & 380    & 0.89$^{+0.05}_{-0.04}$ & 0.87$^{+0.01}_{-0.01}$ & G    \\ \hline
\end{tabular}\label{Table1}
\end{table*}

\begin{table*}[h]\caption{Table \ref{Table1}, continued}
\renewcommand{\arraystretch}{1.5}
\begin{tabular}{|l|l|l|l|l|l|l|}
\hline
Name             & Mass(M$_\oplus$)       & Radius(R$_\oplus$)    & T$_{eq}$ (K)            & Host Mass(M$_\odot$)  & Host Radius(R$_\odot$) & Host type \\ \hline
Kepler-87 c      &  6.40$^{+0.80}_{-0.80}$ & 6.14$^{+0.29}_{-0.29}$ & 403    & 1.10$^{+0.05}_{-0.05}$  & 1.82$^{+0.04}_{-0.04}$ & G    \\ \hline
Kepler-93 b      &  4.54$^{+0.85}_{-0.85}$ & 1.57$^{+0.11}_{-0.11}$ & 1134   & 1.09$^{+0.14}_{-0.14}$ & 0.98$^{+0.04}_{-0.04}$ & G    \\ \hline
TOI-125 b        &  9.50$^{+0.88}_{-0.88}$ & 2.73$^{+0.07}_{-0.07}$ & 1037   & 0.86$^{+0.04}_{-0.04}$ & 0.85$^{+0.01}_{-0.01}$ & G    \\ \hline
TOI-125 c        &  6.63$^{+0.99}_{-0.99}$ & 2.76$^{+0.10}_{-0.10}$ & 828    & 0.86$^{+0.04}_{-0.04}$ & 0.85$^{+0.01}_{-0.01}$ & G    \\ \hline
TOI-125 d        & 13.60$^{+1.20}_{-1.20}$ & 2.93$^{+0.17}_{-0.17}$ & 638    & 0.86$^{+0.04}_{-0.04}$ & 0.85$^{+0.01}_{-0.01}$ & G    \\ \hline
TOI-421 b        &  7.17$^{+0.66}_{-0.66}$ & 2.68$^{+0.19}_{-0.18}$ & 981    & 0.85$^{+0.03}_{-0.02}$ & 0.87$^{+0.01}_{-0.01}$ & G    \\ \hline
TOI-421 c        & 16.42$^{+1.06}_{-1.04}$ & 5.09$^{+0.16}_{-0.15}$ & 674    & 0.85$^{+0.03}_{-0.02}$ & 0.87$^{+0.01}_{-0.01}$ & G    \\ \hline
TOI-561 c        &  5.40$^{+0.98}_{-0.98}$ & 2.88$^{+0.10}_{-0.10}$ & 821    & 0.79$^{+0.02}_{-0.02}$ & 0.85$^{+0.01}_{-0.01}$ & G    \\ \hline
TOI-561 d        & 11.95$^{+1.28}_{-1.28}$ & 2.53$^{+0.13}_{-0.13}$ & 615    & 0.79$^{+0.02}_{-0.02}$ & 0.85$^{+0.01}_{-0.01}$ & G    \\ \hline
TOI-561 e        & 16.00$^{+2.30}_{-2.30}$ & 2.67$^{+0.11}_{-0.11}$ & 426    & 0.79$^{+0.02}_{-0.02}$ & 0.85$^{+0.01}_{-0.01}$ & G    \\ \hline
TOI-763 b        &  9.79$^{+0.78}_{-0.78}$ & 2.28$^{+0.11}_{-0.11}$ & 1038   & 0.92$^{+0.03}_{-0.03}$ & 0.90$^{+0.01}_{-0.01}$  & G    \\ \hline
TOI-763 c        &  9.32$^{+1.02}_{-1.02}$ & 2.63$^{+0.12}_{-0.12}$ & 800    & 0.92$^{+0.03}_{-0.03}$ & 0.90$^{+0.01}_{-0.01}$  & G    \\ \hline
55 Cnc e     &  7.99$^{+0.32}_{-0.33}$ & 1.88$^{+0.03}_{-0.03}$ & 1949   & 0.91$^{+0.01}_{-0.01}$ & 0.94$^{+0.01}_{-0.01}$ & K    \\ \hline
GJ 9827 b    &  4.91$^{+0.49}_{-0.49}$ & 1.58$^{+0.03}_{-0.03}$ & 1184   & 0.61$^{+0.02}_{-0.01}$ & 0.60$^{+0.01}_{-0.00}$      & K    \\ \hline
GJ 9827 d    &  4.04$^{+0.82}_{-0.84}$ & 2.02$^{+0.05}_{-0.04}$ & 686    & 0.61$^{+0.02}_{-0.01}$ & 0.60$^{+0.01}_{-0.00}$      & K    \\ \hline
HD 15337 b   &  7.51$^{+1.09}_{-1.01}$ & 1.64$^{+0.06}_{-0.06}$ & 1001   & 0.90$^{+0.03}_{-0.03}$  & 0.86$^{+0.02}_{-0.02}$ & K    \\ \hline
HD 15337 c   &  8.11$^{+1.82}_{-1.69}$ & 2.39$^{+0.12}_{-0.12}$ & 642    & 0.90$^{+0.03}_{-0.03}$  & 0.86$^{+0.02}_{-0.02}$ & K    \\ \hline
HD 219134 b  &  4.74$^{+0.19}_{-0.19}$ & 1.60$^{+0.06}_{-0.06}$ & 1015   & 0.81$^{+0.03}_{-0.03}$ & 0.78$^{+0.01}_{-0.01}$ & K    \\ \hline
HD 219134 c  &  4.36$^{+0.22}_{-0.22}$ & 1.51$^{+0.05}_{-0.05}$ & 782    & 0.81$^{+0.03}_{-0.03}$ & 0.78$^{+0.01}_{-0.01}$ & K    \\ \hline
HD 219134 d  & 16.17$^{+0.64}_{-0.64}$ & 1.61$^{+0.02}_{-0.02}$ & 410    & 0.81$^{+0.03}_{-0.03}$ & 0.78$^{+0.01}_{-0.01}$ & K    \\ \hline
HD 219134 f  &  7.30$^{+0.40}_{-0.40}$ & 1.31$^{+0.02}_{-0.02}$ & 523    & 0.81$^{+0.03}_{-0.03}$ & 0.78$^{+0.01}_{-0.01}$ & K    \\ \hline
HD 97658 b   &  8.88$^{+1.23}_{-1.23}$ & 2.35$^{+0.11}_{-0.11}$ & 741    & 0.89$^{+0.15}_{-0.15}$ & 0.74$^{+0.01}_{-0.01}$ & K    \\ \hline
HIP 116454 b & 12.71$^{+3.18}_{-3.18}$ & 2.47$^{+0.11}_{-0.11}$ & 740    & 0.81$^{+0.11}_{-0.11}$ & 0.72$^{+0.02}_{-0.02}$ & K    \\ \hline
K2-216 b     &  8.00$^{+1.60}_{-1.60}$ & 1.75$^{+0.17}_{-0.10}$ & 1103   & 0.70$^{+0.03}_{-0.03}$  & 0.72$^{+0.03}_{-0.03}$ & K    \\ \hline
K2-266 e     & 14.30$^{+6.40}_{-5.00}$ & 2.73$^{+0.14}_{-0.11}$ & 490    & 0.69$^{+0.03}_{-0.03}$ & 0.70$^{+0.02}_{-0.02}$  & K    \\ \hline
K2-285 b     &  9.68$^{+1.21}_{-1.37}$ & 2.59$^{+0.06}_{-0.06}$ & 1089   & 0.83$^{+0.02}_{-0.02}$ & 0.79$^{+0.02}_{-0.02}$ & K    \\ \hline
K2-285 c     & 15.68$^{+2.28}_{-2.13}$ & 3.53$^{+0.08}_{-0.08}$ & 741    & 0.83$^{+0.02}_{-0.02}$ & 0.79$^{+0.02}_{-0.02}$ & K    \\ \hline
K2-36 b      &  3.90$^{+1.10}_{-1.10}$ & 1.43$^{+0.08}_{-0.08}$ & 1224   & 0.79$^{+0.01}_{-0.01}$ & 0.72$^{+0.01}_{-0.01}$ & K    \\ \hline
K2-36 c      &  7.80$^{+2.30}_{-2.30}$ & 3.20$^{+0.30}_{-0.30}$ & 788    & 0.79$^{+0.01}_{-0.01}$ & 0.72$^{+0.01}_{-0.01}$ & K    \\ \hline
Kepler-80 b  &  6.93$^{+1.05}_{-0.70}$ & 2.67$^{+0.10}_{-0.10}$ & 709    & 0.73$^{+0.03}_{-0.03}$ & 0.68$^{+0.02}_{-0.02}$ & K    \\ \hline
Kepler-80 c  &  6.74$^{+1.23}_{-0.86}$ & 2.74$^{+0.12}_{-0.10}$ & 641    & 0.73$^{+0.03}_{-0.03}$ & 0.68$^{+0.02}_{-0.02}$ & K    \\ \hline
Kepler-80 d  &  6.75$^{+0.69}_{-0.51}$ & 1.53$^{+0.09}_{-0.07}$ & 936    & 0.73$^{+0.03}_{-0.03}$ & 0.68$^{+0.02}_{-0.02}$ & K    \\ \hline
Kepler-80 e  &  4.13$^{+0.81}_{-0.95}$ & 1.60$^{+0.08}_{-0.07}$ & 815    & 0.73$^{+0.03}_{-0.03}$ & 0.68$^{+0.02}_{-0.02}$ & K    \\ \hline
L 168-9 b    &  4.60$^{+0.56}_{-0.56}$ & 1.39$^{+0.09}_{-0.09}$ & 965    & 0.62$^{+0.03}_{-0.03}$ & 0.60$^{+0.02}_{-0.02}$  & K    \\ \hline
TOI-1235 b   &  6.91$^{+0.75}_{-0.85}$ & 1.74$^{+0.09}_{-0.08}$ & 754    & 0.64$^{+0.02}_{-0.02}$ & 0.63$^{+0.01}_{-0.01}$ & K    \\ \hline
TOI-178 b    &  1.50$^{+0.39}_{-0.44}$ & 1.15$^{+0.07}_{-0.07}$ & 1040   & 0.65$^{+0.03}_{-0.03}$ & 0.65$^{+0.01}_{-0.01}$ & K    \\ \hline
TOI-178 c    &  4.77$^{+0.55}_{-0.68}$ & 1.67$^{+0.11}_{-0.10}$ & 873    & 0.65$^{+0.03}_{-0.03}$ & 0.65$^{+0.01}_{-0.01}$ & K    \\ \hline
TOI-178 d    &  3.01$^{+0.80}_{-1.03}$ & 2.57$^{+0.07}_{-0.08}$ & 690    & 0.65$^{+0.03}_{-0.03}$ & 0.65$^{+0.01}_{-0.01}$ & K    \\ \hline
\end{tabular}
\end{table*}

\begin{table*}[h]\caption{Table \ref{Table1}, continued}
\renewcommand{\arraystretch}{1.5}
\begin{tabular}{|l|l|l|l|l|l|l|}
\hline
Name             & Mass(M$_\oplus$)       & Radius(R$_\oplus$)    & T$_{eq}$ (K)            & Host Mass(M$_\odot$)  & Host Radius(R$_\odot$) & Host type \\ \hline
TOI-178 e    &  3.86$^{+1.25}_{-0.94}$ & 2.21$^{+0.09}_{-0.09}$ & 600    & 0.65$^{+0.03}_{-0.03}$ & 0.65$^{+0.01}_{-0.01}$ & K    \\ \hline
TOI-178 f    &  7.72$^{+1.67}_{-1.52}$ & 2.29$^{+0.11}_{-0.11}$ & 521    & 0.65$^{+0.03}_{-0.03}$ & 0.65$^{+0.01}_{-0.01}$ & K    \\ \hline
TOI-178 g    &  3.94$^{+1.31}_{-1.62}$ & 2.87$^{+0.14}_{-0.13}$ & 470    & 0.65$^{+0.03}_{-0.03}$ & 0.65$^{+0.01}_{-0.01}$ & K    \\ \hline
TOI-776 b    &  4.00$^{+0.90}_{-0.90}$ & 1.85$^{+0.13}_{-0.13}$ & 514    & 0.54$^{+0.03}_{-0.03}$ & 0.54$^{+0.02}_{-0.02}$ & K    \\ \hline
TOI-776 c    &  5.30$^{+1.80}_{-1.80}$ & 2.02$^{+0.14}_{-0.14}$ & 415    & 0.54$^{+0.03}_{-0.03}$ & 0.54$^{+0.02}_{-0.02}$ & K    \\ \hline
TOI-824 b    & 18.47$^{+1.84}_{-1.88}$ & 2.93$^{+0.20}_{-0.19}$ & 1253   & 0.71$^{+0.03}_{-0.03}$ & 0.69$^{+0.03}_{-0.03}$ & K    \\ \hline
\end{tabular}
\end{table*}



\end{appendix}

\end{document}